\documentclass[aps,superscriptaddress]{revtex4}
\usepackage{amssymb}
\usepackage{amsmath}

\usepackage{graphicx}
\usepackage{dcolumn}

\def\p {\partial}
\def\t {\tilde}
\def\be {\begin{equation}}
\def\ee  {\end{equation}}
\def\bea {\begin{eqnarray}}
\def\eea {\end{eqnarray}}
\def\nn {\nonumber}

\begin{document}


\title{Gravitational collapse in quantum gravity}

\author{Viqar Husain}\email{husain@math.unb.ca}
\affiliation{Department of Mathematics and Statistics,\\
University of New Brunswick, Fredericton, NB E3B 5A3, Canada}
  

\thispagestyle{empty}

\date{June 2007}

\begin{abstract}
  
We give a review of recent work aimed at understanding the
dynamics of gravitational collapse in quantum gravity. Its goal is
to provide a non-perturbative computational framework for
understanding the emergence of the semi-classical approximation
and Hawking radiation. The model studied is the gravity-scalar
field theory in spherical symmetry. A quantization of this theory
is given in which operators corresponding to null expansions and
curvature are well defined. Together with the Hamiltonian, these
operators allow one to follow the evolution of an initial
matter-geometry state to a trapped configuration and beyond, in a
singularity free and unitary setting.
  
\end{abstract}


\maketitle



\section{Introduction}

One of the outstanding problems in theoretical physics is the
incomplete understanding at the quantum level of the formation,
and subsequent evolution of black holes in a quantum theory of
gravity. Although a subject of study for over three decades, it is
fair to say that, in spite of partial results in string theory and
loop quantum gravity, there is no widely accepted answer to many
of the puzzles of black hole physics. This is largely because
there has been no study of quantum dynamical collapse in these
approaches. Rather, progress has focused mainly on explanations of
the microscopic origin of the entropy of  static black holes from
state counting. A four-dimensional spacetime picture of black hole
formation from matter collapse, and its subsequent evolution is
not available in any approach to quantum gravity at the present
time.

This paper summarizes an attempt to address this problem in the
context of Hawking's original derivation of black hole radiation:
spherically symmetric gravity minimally coupled to a massless
scalar field. This is a non-linear 2d field theory describing the
coupled system of the metric and scalar field degrees of freedom.
Gravitational collapse in the classical theory in this model has
been carefully studied numerically \cite{chop,gundrev}, but its
full quantization has never been addressed.

Hawking's semi-classical calculation \cite{hawk1} uses the eikonal
approximation for the wave equation in a mildly dynamical
background, where the dynamics centers on the surface of a star
undergoing collapse. The essential content of it is the extraction
of the phase of the ingoing mode from an outgoing solution of the
scalar wave equation as a classically collapsing star crosses its
Schwarzschild radius.  According to this calculation, emitted
particles appear to originate near the event horizon. This means
that an emitted particle observed by a geodesic observer is
transplankian at creation origin due to the gravitational redshift
(which is infinite at the horizon). Its back reaction is therefore
not negligible, bringing into question the entire approximation.

It is likely that a complete understanding of quantum evolution in
this system will resolve all the outstanding problems of black
hole physics in the setting in which they originally arose. The
following sections contain a summary of the work described in
refs. \cite{hw1,hw2,hw3,hw4}.

\section{Classical theory}

The phase space of the model is defined by prescribing a form of
the gravitational phase space variables $q_{ab}$ and
$\tilde{\pi}^{ab}$, together with falloff conditions in $r$ for
these variables, and for the lapse and shift functions $N$ and
$N^a$, such that the ADM 3+1 action for general relativity
minimally coupled to a massless scalar field
\be S = \frac{1}{8\pi G}\int d^3x dt\left[
\tilde{\pi}^{ab}\dot{q}_{ab} + \t{P}_\phi\dot{\phi} - N{\cal H} -
N^a {\cal C}_a\right] \label{act}
\ee
is well defined. The constraints arising from varying the lapse
and shift are
\bea {\cal H} &=&
\frac{1}{\sqrt{q}}\left(\tilde{\pi}^{ab}\tilde{\pi}_{ab}
-\frac{1}{2} \tilde{\pi}^2 \right)-
                       \sqrt{q}\ {\cal R}(q)\nn \\
           &&+ 4\pi G \left(\frac{1}{\sqrt{q}}\tilde{P}_\phi^2
                       + \sqrt{q}q^{ab}\p_a\phi\p_b\phi\right) = 0,  \\
{\cal C}_a &=& D_c\t{\pi}^c_a - \t{P}_\phi\p_a\phi =0,
\eea
where $\t{\pi}=\t{\pi}^{ab}q_{ab}$ and ${\cal R}$ is the Ricci
scalar of $q_{ab}$. The falloff conditions imposed on the phase
space variables are motivated by the Schwarzschild solution in
Painleve-Gullstand (PG) coordinates, which itself is to be a
solution in the prescribed class of spacetimes. These conditions
give the following falloff for the gravitational phase space
variables (for $\epsilon>0$)
\be
q_{ab} = e_{ab} + \frac{f_{ab}(\theta,\phi)}{r^{3/2+\epsilon}}
+ {\cal O}(r^{-2}), \ \ \ \ \ \ \ \ \ \ \pi^{ab} =
\frac{g^{ab}(\theta,\phi)}{r^{3/2}} +
\frac{h^{ab}(\theta,\phi)}{r^{3/2+\epsilon}} + {\cal O}(r^{-2}),
\label{pgfo}
\ee
where $f^{ab},g^{ab},h^{ab}$ are symmetric tensors,
$\pi^{ab}=\t\pi^{ab}/\sqrt{q}$, and $q={\rm det}q_{ab}$.

In this general setting we use the parametrization
\bea
q_{ab} &=& \Lambda(r,t)^2\ n_a n_b + \frac{R(r,t)^2}{r^2}\ ( e_{ab} - n_a n_b)\\
\t{\pi}^{ab} &=& \frac{P_\Lambda(r,t)}{2\Lambda(r,t)}\ n^an^b +
\frac{r^2 P_R(r,t)}{4R(r,t)}\ (e^{ab} - n^a n^b), \label{reduc}
\eea
for the 3-metric and conjugate momentum for a reduction to
spherical symmetry, where $e_{ab}$ is the flat 3-metric and $n^a =
x^a/r$. Substituting these into the 3+1 ADM  action for general
relativity shows that the pairs $(R,P_R)$ and
$(\Lambda,P_\Lambda)$ are canonically conjugate variables. We note
for example the Poisson bracket
\be
 \left\{R_f, e^{i\lambda P_R(r)}\right\}\equiv \left\{ \int_0^\infty Rf\ dr, e^{i\lambda P_R(r)} \right\}
 = i2G\lambda f(r) e^{i\lambda P_R(r)},
\label{bpb}
\ee
which is the bracket represented in the quantum theory (described
below).

The falloff conditions induced on these variables from
(\ref{pgfo}), together with those on the lapse and shift
functions, ensure that the reduced   action
\be
S_R = \frac{1}{2G}\int dtdr \left(P_R\dot{R} +
P_\Lambda\dot{\Lambda} + P_\phi\dot{\phi}
    - {\rm constraint\ terms} \right) + \int_\infty dt (N^r\Lambda P_\Lambda)
\ee
is well defined. This completes the definition of the classical
theory.

At this stage we perform a time gauge fixing using the condition
$\Lambda =1$ motivated by PG coordinates. This is second class
with the Hamiltonian constraint, which therefore must be imposed
strongly and solved for the conjugate momentum $P_\Lambda$. This
gauge fixing eliminates the dynamical pair $(\Lambda,P_\Lambda)$,
fixes the lapse as a function of the shift, and leads to a system
describing the dynamics of the variables $(R,P_R)$ and
$(\phi,P_\phi)$ \cite{hw2}. The reduced radial diffeomorphism
generator
\be C_{red}\equiv P_\Lambda'(R,\phi,P_R,P_\phi) + P_\phi\phi'
+P_RR'=0 \label{reddiff}
\ee
remains as the only first class constraint. It also gives the
gauge fixed Einstein evolution equations via Poisson brackets, for
example $\dot{\phi} = \{\phi, \int dr\ N^r C_{red}\}$.

\section{Quantum gravity}
The quantization route we follow is unconventional in that field
momenta are not represented as self-adjoint operators; rather only
exponentials of momenta are realized on the Hilbert space. This is
similar to what happens in a lattice quantization, except that, as
we see below, every quantum state represents a lattice sampling of
field excitations, with all lattices allowed. This  quantization
allows definitions of bounded inverse configurations operators
such as $1/x$, which for quantum gravity leads to the mechanism
for curvature singularity resolution described below.

A quantum field is characterized by its excitations at a given set
of points in space. The important difference from standard quantum
field theory is that in the representation we use, such states are
normalizable.  A basis state is
\be
|e^{i \sum_k a_k P_R(x_k)}, e^{i L^2\sum_l b_l
P_\phi(y_l)}\rangle \equiv |a_1\ldots a_{N_1};b_1\ldots
b_{N_2}\rangle, \label{basis}
\ee
where the factors of $L$ in the exponents reflect the length
dimensions of the respective field variables, and $a_k,b_l$ are
real numbers which represent the excitations of the scalar quantum
fields $R$ and $\phi$ at the radial locations $\{x_k\}$ and
$\{y_l\}$. The inner product on this basis is
\be
\langle a_1 \ldots a_{N_1};b_1,\ldots b_{N_2}|a'_1 \ldots
a'_{N_1}; b'_1\ldots b'_{N_2} \rangle
 = \delta_{a_1,a_1'}\ldots \delta_{b_{N_2},b_{N_2}'},\nn
\ee
if the states contain the same number of sampled points, and is
zero otherwise.

The action of the basic operators are given by
\bea
&&\hat{R}_f\ |a_1 \ldots a_{N_1};b_1 \ldots b_{N_2}\rangle =
 L^2 \sum_k a_k f(x_k)|a_1 \ldots a_{N_1};b_1 \ldots b_{N_2}\rangle,\\
&& \widehat{e^{i \lambda_j P_R(x_j)}}|a_1\ldots a_{N_1};b_1\ldots
b_{N_2}\rangle
 =  |a_1\ldots, a_j-\lambda_j,\ldots a_{N_1};b_1\ldots b_{N_2}\rangle,
\eea
where $a_j$ is $0$ if the point $x_j$ is not part of the original
basis state. In this case the action creates a new excitation at
the point $x_j$ with value $-\lambda_j$. These definitions give
the commutator
\be
\left[\hat{R}_f,\widehat{e^{i\lambda P_R(x)}} \right] =
-\lambda f(x) L^2 \widehat{e^{i\lambda P_R(x)}}.
\ee
Comparing this with (\ref{bpb}), and using the Poisson bracket
commutator correspondence $\{\ ,\ \}\leftrightarrow i\hbar[\ ,\ ]$
gives $L = \sqrt{2} l_P$, where $l_P$ is the Planck length. There
are similar operator definitions for the canonical pair
$(\phi,P_\phi)$.

\subsection{Singularity resolution}
To address the singularity avoidance issue, we first extend the
manifold on which the fields $R$ etc. live to include the point
$r=0$ , which in the gauge fixed theory is the classical
singularity. We then ask what classical phase space observables
capture curvature information. For homogeneous cosmological
models, a natural choice is the inverse scale factor $a(t)$. For
the present case, a guide is provided by the gauge fixed theory
without matter where it is evident that it is the extrinsic
curvature that diverges at $r=0$, which is the Schwarzschild
singularity. This suggests, in analogy with the inverse scale
factor, that we consider the field variable $1/R$ as a measure of
curvature. A more natural choice would be a scalar constructed
from the phase space variables by contraction of tensors. A simple
possibility is
\be
\t{\pi} = \frac{1}{2}\left(\frac{P_\Lambda}{R^2} +
\frac{P_R}{\Lambda R} \right).
\ee
The small $r$ behaviour of the phase space variables ensures that
any divergence in $\t{\pi}$ is due to the $1/R$ factor. We
therefore focus on this. A first observation is that the
configuration variables $R(r,t)$ and $\phi(r,t)$ defined at a
single point do not have well defined operator realizations.
Therefore we are forced to consider phase space functions
integrated over (at least a part of) space. A functional such as
\be R_f=\int_0^\infty dr fR \ee for a test function $f$ provides a
measure of sphere size in our parametrization of the metric. We
are interested in the reciprocal of this for a measure of
curvature. Since $R\sim r$ asymptotically, the functions $f$ must
have the falloff $f(r) \sim r^{-2-\epsilon}$ for $R_f$ to be well
defined. Using this, it is straightforward to see that $1/R_f$
diverges classically for small spheres: we can choose $f>0$ of the
form $f\sim 1$ for $r << 1$, which for large $r$ falls
asymptotically to zero. Then $R_f\sim r^2$ and $1/R_f$ diverges
classically for small spheres.

A question for the quantum theory is whether $1/R_f$ can be
represented densely on a Hilbert space as $1/\hat{R}_f$.  This is
possible only if the chosen representation is such that
$\hat{R}_f$ does not have a zero eigenvalue. If it does, we must
represent $1/R_f$ as an operator more indirectly, using another
classically equivalent function. Examples of such functions are
provided by Poisson bracket identities such as
\be
\frac{1}{|R_f|} = \left(\frac{2}{iG f(r)}\
e^{-iP_R(r)}\left\{\sqrt{|R_f|}, e^{iP_R(r)} \right\}\right)^2,
\label{invRf}
\ee
where the functions $f$ do not have zeroes. The representation for
the quantum theory described above is such that the operator
corresponding to $R_f$ has a zero eigenvalue. Therefore we
represent $1/R_f$ using the r.h.s. of (\ref{invRf}). The central
question for singularity resolution is whether the corresponding
operator is densely defined and bounded. This turns out to be the
case.

Using the expressions for the basic field operators, we can
construct an operator corresponding to a classical singularity
indicator:
\be
\widehat{\frac{1}{|R_f|}} \equiv
  \left( \frac{2}{l_P^2 f(x_j)}\widehat{e^{- iP_R(x_j)}}\left[ \widehat{\sqrt{|R_f|}},
\ \widehat{e^{iP_R(x_j)}} \right] \right)^2. \label{1/R}
\ee
The result is that basis states are eigenvectors of this operator,
and all eigenvalues are bounded. This is illustrated   with the
state
\be
|{\rm S}_{a_0}\rangle \equiv|e^{ia_0P_R(r=0)}\rangle,
\ee
which represents an excitation $a_0$ of the quantum field
$\hat{R}_f$ at the point of the classical singularity:
\bea
\hat{R}_f |{\rm S}_{a_0}\rangle &=& (2 l_P^2) f(0) a_0\ |{\rm S}_{a_0}\rangle, \\
\widehat{\frac{1}{|R_f|}}|{\rm S}_{a_0}\rangle  &=& \frac{2}{l_P^2
f(0)}\left(|a_0|^{1/2} - |a_0-1|^{1/2}\right)^2 |{\rm
S}_{a_0}\rangle  \nn
\eea
which is clearly bounded. This shows that the singularity is
resolved at the quantum level. In particular if there is no
excitation of $R_f$ at the classical singularity, ie. $a_0=0$, the
upper bound on the eigenvalue of the inverse operator is
$2/l_P^2$.
\subsection{Quantum black holes}
The event horizon of a static or stationary black hole is a global
spacetime concept. It does not provide a useful local
determination of whether one is inside a black hole. The
fundamental idea for defining a black hole locally is that of a
trapped surface, first introduced by Penrose. One considers a
closed spacelike 2-surface in a spacetime, and computes the
expansions $\theta_+$ and $\theta_-$ of outgoing and ingoing null
geodesics emanating orthogonally from the surface. If $\theta_+
>0$ and $\theta_- <0$, the surface is considered normal. On the
other hand if $\theta_+\le 0$ and $\theta_- <0$, the surface is
called trapped. This provides a criterion for subdividing a
spacetime into trapped and normal regions.  The outer boundary of
a trapped region may be considered as the (dynamical) boundary of
black hole, also known as the "apparent horizon" in numerical
relativity. It is a function computed in classical numerical
evolutions to test for black hole formation. Similarly, a setting
for studying quantum collapse requires an operator realisation of
the null  expansion "observable,"  and a criterion to see if a
given quantum state describes a "quantum black hole."

The classical expansions in spherical symmetry are the phase space
functions \cite{hw3}
\be
\theta_\pm = -\frac{1}{2\Lambda}\left(2R^2\Lambda \Lambda' \pm
P_\Lambda + 4\Lambda^2 RR'\right). \label{hor}
\ee
Given phase space functions on a spatial hypersurface $\Sigma$,
the marginal trapping horizon(s) are located by finding the
solution coordinates $r=r_i$ ($i=1\cdots n$) of the conditions
$\theta_+=0$ and $\theta_-<0$, (since in general there may be more
than one solution). The corresponding radii $R_i= R(r_i)$ are then
computed. The size of the horizon on the slice $\Sigma$ is the
largest value in the set $\{R_i\}$.

Since only translation operators are available in our
quantization, we define $P_\Lambda$ indirectly by
\be
  \hat{P}_\Lambda^\lambda = \frac{l_P}{2i\lambda}\ \left( \hat{U}_\lambda - \hat{U}_\lambda^\dagger\right)
\ee
where $0<\lambda\ll 1$ is an arbitrary but fixed parameter, and
$U_\lambda$ denotes exp$(i\lambda P_\Lambda/L)$. This is motivated
by the corresponding classical expression, where the limit
$\lambda\rightarrow 0$ exists, and gives the classical function
$P_\Lambda$. $\lambda$ is perhaps best understood as a ratio of
two scales, $\lambda = l_p/l_0$, where  $l_0$ is a system size. As
for a lattice quantisation, it is evident that momentum in this
quantisation can be given approximate meaning only for $\lambda\ll
1$. $\lambda$ is also the minimum value by which an excitation can
be changed.

Definitions for the operators corresponding to $R'$ and $\Lambda'$
are obtained by implementing the idea of finite differencing.  We
use narrow Gaussian smearing functions with variance proportional
to the Planck scale, peaked at coordinate points $r_k+\epsilon
l_P$, where $0<\epsilon \ll 1$ is a parameter designed to sample
neighbouring points:
\be
f_{\epsilon}(r,r_k)= \frac{1}{\sqrt{2\pi}}\ {\rm exp} \left[-\
\frac{(r-r_k-\epsilon l_P)^2} {2 l_P^2}\right]
\ee
Denoting $R_{f_\epsilon}$ by $R_{\epsilon}$ for this class of test
functions we define
\be
  \hat{R'}(r_k):= \frac{1}{l_P \epsilon}\ \left( \hat{R}_{\epsilon} - \hat{R}_{0} \right).
\label{R'}
\ee
Putting all these pieces together, we can construct the desired
operators
\be
\hat{\theta}_\pm(r_k) = -\frac{2}{\epsilon l_P} \hat{R}_0^2
\hat{\Lambda}_0^2 \left(\hat{\Lambda}_\epsilon - \hat{\Lambda}_0
\right) \mp \frac{l_P}{2i\lambda}\left( \hat{U}_\lambda -
\hat{U}_\lambda^\dagger \right)\hat{\Lambda}_0 - \frac{4}{\epsilon
l_P}\hat{\Lambda}_0^3 \hat{R}_0 \left(\hat{R}_\epsilon - \hat{R}_0
\right),
\ee
which have a well defined action on the basis states.

In analogy with the classical case, we propose that {\it a state
$|\Psi\rangle$ represents a quantum black hole} if
\be
\langle \Psi| \hat{\theta}_+(r_k) |\Psi\rangle = 0, \ \ \ {\rm
and}\ \ \ \langle \Psi| \hat{\theta}_-(r_k) |\Psi\rangle <0.
\label{qbh}
\ee
for some $r_k$. The corresponding {\it horizon size} is given by
$R_H = \langle \Psi| \hat{R}(r_k) | \Psi\rangle$.

This definition is utilised as follows: Given a state with field
excitations at a set of coordinate points $\{r_i\}$, one would
plot the expectation values in Eqn. (\ref{qbh}) as functions of
$\langle \Psi| \hat{R}(r_k) |\Psi\rangle$, and locate the zeroes,
if any, of the resulting graph.  The resulting "quantum horizon"
location is invariant under radial diffeomorphisms because these
act on states to shift the coordinate locations of field
excitations, but leave the expectation values unchanged -- the
graph is a physical observable.

It is straightforward to construct explicit examples of states
satisfying  these quantum trapping conditions. Some examples
appear in \cite{hw3}. The quantum horizons so determined are not
sharp since $\langle \theta_+^2 \rangle \ne 0$.
\section{Summary and outlook}
The results so far from this approach to understanding black hole
formation in quantum gravity are threefold: (i) A quantization
procedure which allows explicit calculations to be done, (ii) a
test for black holes in a full quantum gravity setting which makes
no use classical boundary conditions at event horizons, and (iii)
singularity free and unitary evolution equations using the
Hamiltonian defined in \cite{hw4}.

The main computational challenge is to use the formalism to
explicitly compute the evolution of a given matter-geometry state
until it satisfies the quantum black hole criteria, and then to
continue to the evolution to see if and how Hawking radiation
might arise. This work is in progress.

\end{document}